\documentclass{article}

\usepackage{hyperref}
\usepackage{graphics,graphicx,url,verbatim,xspace}
\usepackage{bold-extra}
\usepackage[usenames, dvipsnames]{color}
\usepackage{enumitem}
\usepackage{natbib}
\usepackage{lipsum}
\usepackage{eurosym}
\usepackage{fancyhdr}
\usepackage{color}
\definecolor{black}{rgb}{0,0,0}
\definecolor{red}{rgb}{1.0,0,0}

\usepackage{graphicx}
\usepackage{subfig}
\usepackage{authblk}

\newcommand{\BL}{\textit{Breakthrough Listen}\xspace}

\definecolor{black}{rgb}{0,0,0}
\definecolor{red}{rgb}{1.0,0,0}

\title{\BL Follow-up of the Reported Transient Signal Observed at the Arecibo Telescope in the Direction of Ross 128}

\author[1,2]{J. Emilio Enriquez} 
\author[1,2,3]{Andrew Siemion}
\author[1]{Ryan Dana} 
\author[1]{Steve Croft}
\author[4]{Abel M\'{e}ndez}
\author[1]{Andrew Xu}
\author[1]{David DeBoer}
\author[5]{Vishal Gajjar} 
\author[1]{Greg Hellbourg}
\author[1]{Howard Isaacson}
\author[1]{Matt Lebofsky}
\author[1]{David H.\ E.\ MacMahon}
\author[1,6]{Danny C.\ Price}
\author[5]{Dan Werthimer}
\author[7]{Jorge Zuluaga}

\affil[1]{Department of Astronomy, University of California, Berkeley, 501 Campbell Hall \#3411, Berkeley, CA, 94720, USA}
\affil[2]{Department of Astrophysics/IMAPP, Radboud University, P.O. Box 9010, NL-6500 GL Nijmegen, The Netherlands} 
\affil[3]{SETI Institute, Mountain View, California}
\affil[4]{PHL, University of Puerto Rico at Arecibo}
\affil[5]{Space Sciences Laboratory, University of California, Berkeley}
\affil[6]{Centre for Astrophysics \& Supercomputing, Swinburne University of Technology, PO Box 218, Hawthorn, VIC 3122, Australia}
\affil[7]{Institute of Physics / FCEN - Universidad de Antioquia}

\begin{document}
\pagestyle{plain}
\pagenumbering{arabic}
\maketitle

\begin{abstract}
We undertook observations with the Green Bank Telescope,  simultaneously with the 300m telescope in Arecibo, as a follow-up of a possible flare of radio emission from Ross 128. We report here the non-detections from the GBT observations in C band (4-8 GHz), as well as non-detections in archival data at L band (1.1-1.9 GHz). We suggest that a likely scenario is that the emission comes from one or more satellites passing through the same region of the sky.  
\end{abstract}

\section{Introduction.}
\label{sec:intro}

Ross 128 is a low mass red dwarf star (M4V) and one of the nearest stars to the solar system at a distance of only 10.9 light-years (3.3pc). It has been known as a flare star in the optical for more than four decades \citep{1972IBVS..707....1L}. 
Nonetheless, we are not aware of any radio detections of Ross 128 presented in the literature, although radio observations yielding upper limits have been reported \citep[][and others]{1989ApJS...71..895W}. It may still be possible that non-thermal radio emission is originating from this star and could even be detectable given the star's close proximity to us. Also, radio flares from other M dwarfs have previosly been seen, the most well known example is AD Leo\citep{2006ApJ...637.1016O}, a young M dwarf at 5pc away.

On 2017 May 12th, \cite{MendezOnline} observed a group of red dwarfs (including Ross 128) with the 300 meter telescope at the Arecibo Observatory (AO), in a search for non-thermal emission possibly associated with exoplanets in these systems.  We note that no exoplanets are known in the Ross 128 system. \cite{MendezOnline} detected anomalous unpolarized radio emission during the observation of Ross 128 lasting for the full 10 min observation and spanning frequencies between 4.6 and 4.8 GHz.  The observed emission has similar morphology and dispersive-like features to those presented by \cite{2008ApJ...674.1078O} on AD Leo, but those emissions were between 1100 and 1600 MHz and were highly circularly polarized.  

\cite{MendezOnline} suggested that the emission was not likely to be local radio frequency interference (RFI) since it was not detected during observations of other stars preceding and following Ross 128.  However, \cite{MendezOnline} was unable to determine conclusively whether the emission seen during the observation of Ross 128 was either associated with the star, was due to another source along the line of sight or was caused by RFI.  The serendipitous potential detection of radio emission from this source encouraged a re-observation by the Mendez group. New observations were carried out on July 16th with Arecibo.   

In order to investigate the possible stellar nature of the Ross 128 emission, the \textit{Breakthrough Listen} Team joined in this campaign and observed Ross 128 simultaneously with Arecibo using the Green Bank Telescope (GBT) and its C-band  receiver (4$-$8 GHz). Here, we briefly report on the observations carried out with the GBT during this campaign. We also present analysis of \BL archival data on this same source using the L-band receiver (1.1-1.9 GHz) of the GBT that has recently been submitted for publication \citep{2017arXiv170903491E}.  The data used for this work, along with plots of spectra across the full GBT bandwidth, is available online\footnote{ \url{seti.berkeley.edu/ross128}}.

\section{BL observations}
\label{sec:obs}

On 2017 July 16, the \BL Team conducted C-band (4--8\,GHz) observations of Ross 128 using the GBT. The current \BL digital back-end \citep{MacMahonBL2017} is capable of operating over the full 4\,GHz span of the receiver simultaneously. The \BL back-end is unique in the sense that it can record time series voltage data across the full band, allowing arbitrary time and frequency resolution depending on science needs. For the case of  C-band observations, the storage needs correspond to 120 Gbps for the raw data.

During the campaign we conducted three five minute observations of Ross 128 interleaved by five minute observations of an OFF source star (Hipparcos 55848). This set of observations was followed by a 15 minute on-source observation of Ross 128. 

Currently three standard data products are produced from each observation \citep{LebofskyBL2017}; one high frequency resolution product primarily for SETI analysis, one high time resolution product for broadband fast transient applications, and one intermediate frequency resolution for other astrophysical radio emission. The high resolution filterbank products are about 50 times smaller than the raw voltage data, and the intermediate resolution products are about 7 times smaller still. 
All the different data products are stored permanently and will be publicly available online.\footnote{\url{https://seti.berkeley.edu/listen/}}.

For this analysis we have used the intermediate frequency resolution product, which has a one second time resolution and three kHz frequency resolution, both well matched to the phenomena under investigation. The bandpass of the C-band and polyphase filter applied to the raw data for channelization was removed by simply dividing by the median value of the time variations for a given frequency channel.

The current \BL SETI search pipeline, as described by \citet{2017arXiv170903491E}, is optimized for the detection of narrow-band Doppler-drifting signals. Since the emission reported by \citet{MendezOnline} is broader and more complex in nature, we perform a by-eye comparison of the ON and OFF observations, in a search for emission uniquely coming from the direction of Ross 128. No such emission is seen in our data.

Based on a conservative estimate of sensitivity from a typical C-band system equivalent flux density of 10\,Jy, we calculate a $3\sigma$ upper limit of $\sim$500\,mJy for impulsive emission at these time and frequency scales. The sensitivity of the GBT at C band is comparable to that of Arecibo. \cite{Mendez2017} reported a 1$\sigma$ sensitivity of 0.12\,Jy for impulsive emission.

\begin{figure}[!htb]
\begin{center}
\includegraphics[width=1.\linewidth]{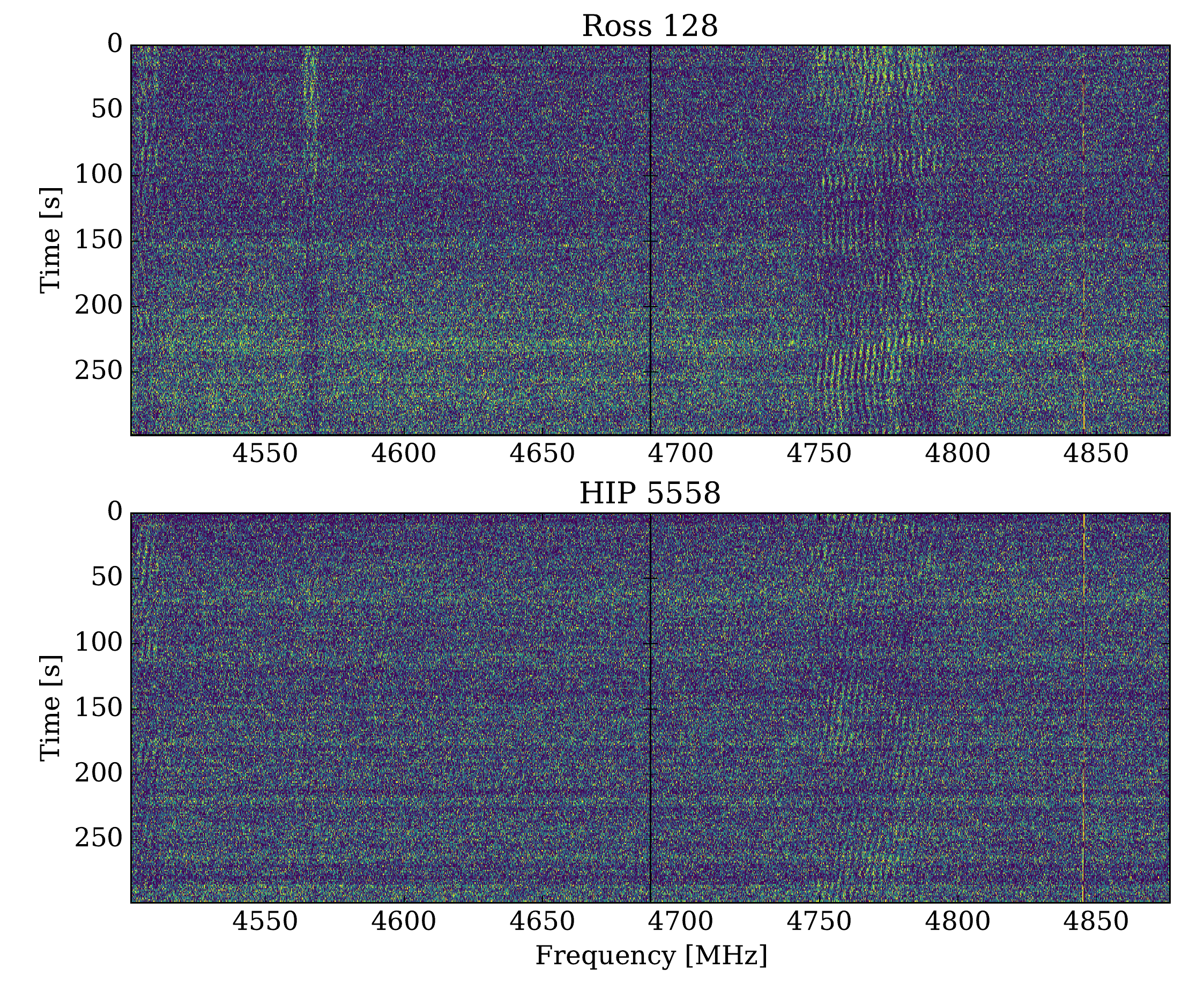}
\caption{Waterfall (time/frequency, Stokes I) plots from representative five minutes observations of Ross 128 and HIP\,55848 from the GBT campaign discussed herein. The two frequency ranges shown correspond to two 187.5 MHz subsets of the total 4 GHz band observed.}
\label{fig:ross128_GBT_C-band}
\end{center}
\end{figure}

Figure \ref{fig:ross128_GBT_C-band} shows a representative example of our observations of Ross 128 and the off-source HIP\,55848. We show here the frequency range between 4.5 to 4.9 GHz which corresponds to part of the region where \cite{Mendez2017} detected emission (4.6-4.8 GHz).  The common spectral features seen on both Ross 128 and HIP\,55848 are examples of interference and instrumental effects.  From our observations it is evident that there is no distinct presence of any emission isolated to Ross 128.   We point out that the morphology of the emission seen betteen 4.75 and 4.8 GHz is similar to that of a Single Sideband Suppressed Carrier \citep{Technical_Handbook}, widely used for communications applications.

\section{Archival data}
\label{sec:arc_data}

Ross 128 was observed on 2016 May 16 with the L-band receiver on GBT as part of the regular \BL targeted search for extraterrestrial artificial radio emission \citep{2017PASP..129e4501I,2017arXiv170903491E}. These observations were carried out using a similar ON-OFF approach as the one used for the C-band observations.

We took the available data at the optimal frequency resolution (same as described here for C band) for this project and inspected these data for broadband emission as well. No such emission was found in our observations at similar sensitivity to the C-band observations.

\section{Discussion and Conclusion}
\label{sec:conclusion}
We see no evidence for any non-thermal radio emission during either our recent GBT C-band observations of Ross 128, or in archival GBT L-band data.  However, our observations were of limited duration and we can not constrain the emission observed in the \citet{MendezOnline} observations to be either astrophysical in nature or RFI.  If Ross 128 does indeed produce detectable non-thermal radio emission it could well be intermittent and require substantial follow-up to confirm.   

\begin{figure}[!htb]
\begin{center}
\begin{tabular}{c}
\includegraphics[width=0.9\linewidth]{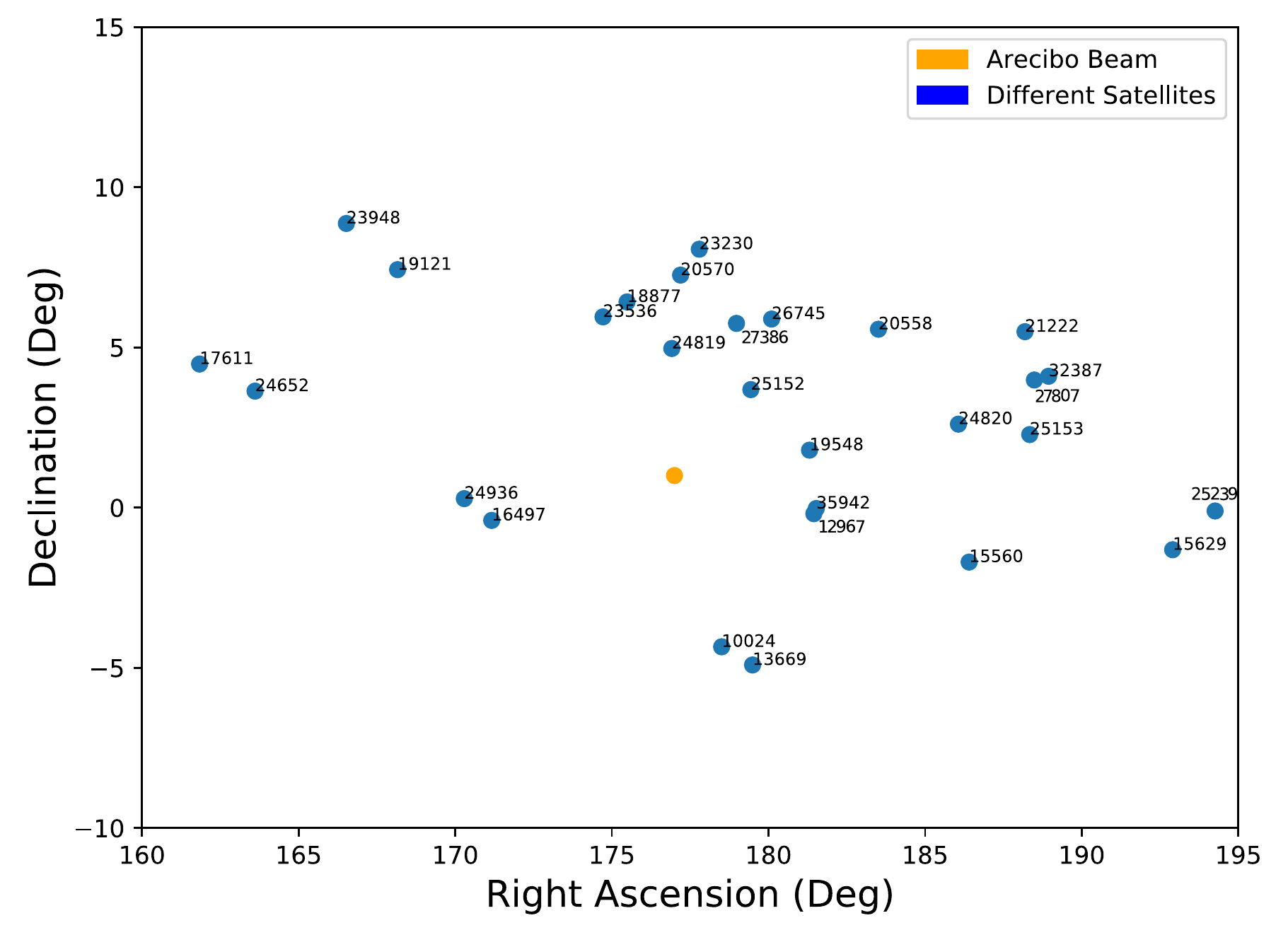}
\end{tabular}
\caption{Position of known satellites with radio transmitters operating between 4-8 GHz. The location of Ross 128 is (ra, dec) $\sim$ (177$^{o}$, 1$^{o}$). Each blue symbol has a Satellite Catalog Number from the NORAD Catalog (also known as the NASA Catalog), information on individual satellites can be found here: \url{https://nssdc.gsfc.nasa.gov/nmc/}}
\label{fig:sats-cband}
\end{center}
\end{figure}

We note that one possible source for the emission detected by \citet{MendezOnline} could be a broadband radio signal from an artificial satellite. Figure \ref{fig:sats-cband} shows the position of geosynchronous satellites around the direction of Ross 128.  The satellite locations were derived from Two Line Element(TLE) \footnote{Taking into account the different values one can accurately predict the RA and Declination of a satellite.}  sources provided by the online space situational awareness (SSA) services, open to the public, provided by the U.S. Strategic Command (USSTRATCOM)\footnote{\url{https://www.space-track.org/}}.  The position of the Arecibo Observatory was used as observing reference. Only those satellites specifically emitting in C-band frequencies are shown. 

We note that the presence of RFI within the same band observed in the \citet{MendezOnline} work shows that this band is (at least at the time of the observations) occupied by interference, and raises the likelihood that the \citeauthor{MendezOnline} emission is indeed due to RFI.

This work shows the potential for using the \textit{Breakthrough Listen} backend as well as existing observations to look for flare emission from nearby stars. Access to the raw voltage time series data enables us to optimize the frequency and time resolution of the filterbank data products to match the needs of the analysis. In future work, we plan to look for flare emission in all the existing \textit{Breakthrough Listen} data.

\section{Acknowledgements}

Breakthrough Listen is managed by the Breakthrough Initiatives, sponsored by the Breakthrough Prize Foundation \footnote{\url{www.breakthroughinitiatives.org}}.

\bibliographystyle{yahapj}
\bibliography{refences}

\end{document}